\documentclass{ws-procs975x65}

\usepackage{amsmath}
\usepackage{amssymb}

\newcommand{\BbbZ}{\mathbb{Z}}

\newcommand{\uone}{\mathrm{U}(1)}

\begin{document}

\title{CHARGED UNRUH EFFECT ON GEON SPACETIMES}
\author{DAVID E. BRUSCHI and JORMA LOUKO}
\address{School of Mathematical Sciences, University of Nottingham, 
\\
Nottingham NG7 2RD, United Kingdom}

%\maketitle

\begin{abstract}
A topological geon black hole with gauge charges 
may have a gauge bundle that necessarily incorporates 
charge conjugation as a gauge symmetry. 
This happens for example for the Reissner-Nordstr\"om geon. 
We show that gauging the charge conjugation leaves 
an imprint in the Unruh effect: 
the geon's exterior region contains 
non-thermal correlations 
for particle pairs of the \emph{same\/}, 
rather than opposite, charge. 
The phenomenon occurs also in topologically similar  
Rindler spacetimes with a background gauge field. 

$\phantom{xxx}$ 

\noindent 
Talk given 
by David E. Bruschi 
at the 
{\sl 12th Marcel Grossmann Meeting\/}, 
Paris, France, 12--18 July 2009. 
\end{abstract}

\bodymatter
%%%%%%%%%%%%%%%%%%%%%%%%%%%%%%%%%%%
\section{Introduction}
%%%%%%%%%%%%%%%%%%%%%%%%%%%%%%%%%%%

Given a stationary black hole spacetime 
with a bifurcate Killing horizon, 
it may be possible to form a time-orientable 
quotient spacetime in which 
the two opposing exterior regions 
have become identified. 
In the Hawking-Unruh effect on the resulting topological 
geon black holes 
\cite{Sorkin:1979ja,friedmansorkin80,friedmansorkin82,sorkin86,%
Friedman:1993ty,Louko:1998hc,Louko:2000tp,%
Maldacena:2001kr,Louko:2004ej}, 
a suitably limited set of observations in the exterior region 
displays thermality in the usual Hawking temperature, 
but there are also non-thermal correlations that bear an 
imprint of the unusual geometry behind the horizons 
\cite{Louko:1998hc,Louko:2000tp,Maldacena:2001kr,%
Louko:2004ej,Louko:1998dj,Langlois:2004fv}. 
In a sense, the Hawking-Unruh effect 
on a geon black hole 
reveals to an exterior observer 
features of the geometry that are 
classically hidden behind the horizons. 
A recent review can be found in~\cite{Louko:2010tq}. 

When the geon has a gauge field, it may be necessary for the 
geon's gauge bundle to be nontrivial in a way that incorporates 
charge conjugation as a gauge symmetry
\cite{Louko:2004ej,Kottanattu:2010zg,Kiskis:1978ed}. 
This happens for example for the
Reissner-Nordstr\"om geon, 
both with electric and magnetic charge \cite{Louko:2004ej}. 
In this contribution we show that 
the Hawking-Unruh effect on the geon bears  
an imprint of the 
gauged charge conjugation: 
the geon's exterior region contains non-thermal correlations 
for particle pairs of the \emph{same\/}, 
rather than opposite, charge. 

We focus this contribution 
on 
two geon-type 
Rindler spacetimes \cite{Louko:1998hc} 
with a background magnetic field. 
The black hole case is 
qualitatively similar \cite{BruschiLouko}.

\section{Magnetic Rindler geons}

Given Minkowski space with global coordinates $(t,x,y,z)$, 
we denote by 
$\mathcal{M}_{0}$ the spacetime 
in which the $z$-coordinate 
is periodic with period~$L>0$. 
We introduce on 
$\mathcal{M}_{0}$ the 
background Maxwell field 
with the globally-defined gauge potential 
\begin{align}
A = - C y \mathrm{d} z
\label{Zwei}
\end{align}
where $C$ is a constant. 

$\mathcal{M}_{0}$ admits the freely-acting involutive isometries 
\begin{align}
J_{\pm}:(t,x,y,z)\longmapsto \bigl(t,-x,\pm y,z+\tfrac12 L\bigr), 
\end{align}
and we denote the respective quotients 
of $\mathcal{M}_{0}$ by~$\mathcal{M}_{\pm}$. 
As the gauge potential 
\eqref{Zwei} is invariant under~$J_+$, 
the gauge bundle over 
$\mathcal{M}_+$ is the trivial $\uone$ bundle. 
However, as the gauge potential 
\eqref{Zwei} changes sign under~$J_-$, 
the gauge bundle over 
$\mathcal{M}_-$ has gauge group 
$\mathrm{O}(2) \simeq \mathbb{Z}_2 \ltimes \mathrm{U}(1)$, 
where the nontrivial element of 
$\mathbb{Z}_2$ acts on $\mathrm{U}(1)$ 
by complex conjugation~\cite{Kiskis:1978ed}. 
The gauge transformation that 
compensates for the minus sign 
in the action of $J_-$ is 
$(-1_{\BbbZ_2}, 1_{\BbbZ_2}): A \mapsto -A$
\cite{Kottanattu:2010zg}. 

$\mathcal{M}_{0}$ is analogous to
magnetic Reissner-Nordstr\"om, with the 
Rindler wedges at $x> |t|$ and $x < - |t|$ 
corresponding to the opposing Reissner-Nordstr\"om exteriors. 
$\mathcal{M}_{\pm}$ are each analogous to the 
Reissner-Nordstr\"om geon, but it is only in 
$\mathcal{M}_{-}$ that the analogue 
extends to the gauged 
charge conjugation in the gauge bundle. 

\section{Charged scalar field on $\mathcal{M}_{\pm}$} 

We consider on $\mathcal{M}_{\pm}$ a complex scalar field $\phi$ 
with the Lagrangian 
\begin{equation}
\mathcal{L}=-(D_{\mu}\phi)^{*}D^{\mu}\phi-m^{2}\phi^{*}\phi
\end{equation}
where the gauge-covariant derivative 
$D_\mu := \nabla_\mu -i e A_\mu$ 
contains the coupling to the background gauge field
and ${}^*$ denotes complex conjugation. 
Note that when the gauge group is 
$\mathbb{Z}_2 \ltimes \mathrm{U}(1)$, 
the action of the disconnected component on  
$\phi$ includes complex conjugation: 
in particular, 
$(-1_{\BbbZ_2}, 1_{\BbbZ_2})$ 
has the action $(A, \phi) \mapsto (-A, \phi^*)$. 
On $\mathcal{M}_{-}$,  
positive and negative charges are thus gauge-equivalent. 

We quantise $\phi$ on $\mathcal{M}_{\pm}$ 
in the usual fashion 
and prepare it in the  
Minkowski-like vacuum $\left|0_{\pm}\right\rangle$, 
defined 
in terms of positive and negative frequencies 
with respect to the Killing vector~$\partial_t$. 
Note that 
$\partial_t$ is globally defined on~$\mathcal{M}_{\pm}$ and
the purely spatial gauge 
potential 
does not affect the positive and negative frequencies. 
On~$\mathcal{M}_{-}$, gauge invariance under 
the disconnected component of the gauge group 
can be handled like the restriction of 
a complex scalar field to 
a real scalar field \cite{BruschiLouko}. 

\section{Unruh effect} 

To analyse the Unruh effect, the task is now to 
express 
$\left|0_{\pm}\right\rangle$
in terms of Rindler 
excitations on the Rindler vacuum. 
We suppress here the formulas \cite{BruschiLouko}
and just describe the qualitative outcome. 

The result has expected similarities to that  
for the real scalar field \cite{Louko:1998dj}. 
In particular, the 
Rindler excitations come (for 
generic quantum numbers) 
in correlated pairs, 
and observations that only see one member of 
each pair are thermal in the usual Unruh temperature. 

The 
new phenomenon is in the charge of the Rindler excitations. 
Observe first that this notion is well defined
not only for 
$\mathcal{M}_{+}$ but also for~$\mathcal{M}_{-}$, 
despite the gauged charge conjugation 
on~$\mathcal{M}_{-}$. 
The reason is that 
$J_-$ maps the two Rindler wedges of 
$\mathcal{M}_0$ to each other, and the 
charge gauging on $\mathcal{M}_{-}$ can thus be 
fixed within the Rindler 
wedge. 

\newpage 

We find: 
\begin{itemize}
\item 
In $\left|0_{+}\right\rangle$ the two Rindler particles 
in each correlated 
pair have the \emph{opposite\/} charge. 
This is as expected: the same can be 
verified to hold also for a complex scalar 
field in the absence of a background gauge field. 
\item 
In $\left|0_{-}\right\rangle$, 
the two Rindler particles in each correlated 
pair have the \emph{same\/} charge. 
This is a direct consequence of the 
gauged charge conjugation in the gauge bundle. 
\end{itemize}

\section{Conclusions}

On a Rindler version of 
the magnetic Reissner-Nordstr\"om geon, 
non-thermal correlations 
in the Unruh effect reveal to a Rindler 
observer that charge conjugation 
has become a gauge symmetry, 
even though the charge gauging  
only affects the gauge 
bundle behind the Rindler horizons. 
The case for the genuine Reissner-Nordstr\"om geon 
is qualitatively similar \cite{BruschiLouko}.

%%%%%%%%%%%%%%%%%%%%%%%%%%%%%%%%%%%

\end{document}